# Is it Possible to Accurately Identify a Shooter's Face as Illuminated by the Muzzle Flash of a 22 LR Pistol?

Michael Courtney,  PhD
BTG Research, Baton Rouge, Louisiana
Michael_Courtney@alum.mit.edu

**Abstract**
The science of firearms muzzle flash has been dominated by three perspectives: 1) Does the muzzle flash reveal friendly positions to naked eye observers so as to draw fire from enemy combatants? 2) Can flash signatures be recognized by electronic surveillance with sufficient accuracy to identify source firearms?  3)  Is the muzzle flash so bright as to diminish the night vision of friendly forces during a firefight?  This paper addresses a fourth question:  Does the muzzle flash of a specific cartridge provide sufficient duration and intensity of illumination with visible light to allow observers to positively identify a shooter's face?  The experiment used two shooters firing a total of 20 shots from a 22 LR semi-automatic pistol.  Shooters fired in randomized order, while eight observers (male US military veterans) attempted to identify the shooter for each trial shot.  The firing range was mostly darkened with only enough ambient light to safely conduct firing tests at the direction of a range safety officer.  There was not enough ambient light to identify the shooter's face.  Observers correctly identified the shooter 54% of the time and failed to correctly identify the shooter 46% of the time.  Observers were unanimous that there was no hint of any visible illumination on the shooter's face provided by the muzzle flash.  In cases where the shooter was correctly identified, observers reported using visual cues such as the sillhouette of the shooters' hair and slight differences in arm positions when firing the pistol.  Other people in the room (including the author) were in unanimous agreement that there was no illumination of the shooters' faces from discharge of a 22 LR pistol.  Based on this experiment, it is concluded with a high degree of scientific certainty that accurate visual identification of a shooter's face is not possible from the illumination provided by a 22 LR muzzle flash with the ammunition used in testing.  It remains an open question whether a shooter's face could be identified using night vision equipment or IR sensitive cameras or electronics.

**Introduction**

Rifles and pistols operate using pressure of hot gases generated by rapid combustion of propellants (predominantly nitrocellulose and nitroglycerin) to accelerate a bullet to high velocities through a rifled barrel.  Once the propellant is ignited, the bullet exits the barrel within 1-2 milliseconds followed by a brief jet of hot expanding gases.  In the same way that all hot matter can produce electromagnetic radiation in the infrared (IR), visible, or ultraviolet (UV) spectrum, hot gases also produce electromagnetic radiation according to their temperature and chemical composition.  Since expanding gases cool very quickly, the duration of infrared radiation produced by hot gases escaping from a firearm barrel is very short (less than 30 milliseconds), and the duration of visible light produced is even shorter.  One study (Burke and Bratlie, 2011) found the duration of muzzle flashes, including IR components (up to 1100 nm), to have durations from 1-2 milliseconds, except for 30-06 cartridges, which occasionally produced flash durations up to 7 milliseconds.

The amount, visibility, and duration of electromagnetic radiation produced by propellant gases escaping from the muzzle of a firearm depends on various factors including the quantity of propellant, presence of flash suppressing compounds, propellant burn rate, ratio of fuel to oxygen in the propellant, projectile speed, and speed of the gases exiting the muzzle. In cases of fuel rich (oxygen deficient) propellants and supersonic projectiles, the intersection of shock fronts of the bullet and propellant gases can generate sufficient heat to re-ignite unburned fuel in the presence of atmospheric oxygen. This phenomenon creates a brighter muzzle flash, but it has not been reported to occur with subsonic projectiles. Being a complex phenomenon with many contributing factors, there seems to be no reliable techniques in the published literature for prediction of intensity, duration, or spectral composition of muzzle flash from a given cartridge. (Burke and Bratlie, 2011) Consequently, investigation into reported muzzle flash of any specific cartridge and ammunition requires an experiment that replicates the desired conditions.

Three main perspectives have dominated research into firearms muzzle flash. The first concern driving muzzle flash research was the propensity for muzzle flash to reveal friendly positions during low light encounters and draw fire from enemy combatants. (Haag, 2007) This concern has led to addition of flash suppressing chemicals to military powders and to flash suppressing muzzle devices attached to barrels. Flash suppressing powders reduce incidence of secondary ignition and tend to shift the radiative energy emitted by escaping gases from the visible to the infrared region of the spectrum, hiding muzzle flash from naked eye observers rather than IR sensitive night vision and surveillance equipment. Muzzle devices tend to disrupt and change the direction of shock waves as they propagate from the muzzle so interacting shock fronts are cooler and less likely to re-ignite gaseous fuels in the presence of atmospheric oxygen.

A second concern of muzzle flash research has been recognizing flash signatures by electronic surveillance to recognize the beginning of attacks, locations of enemy combatants, and identify friend or foe. (Merhav et al., 2013) Electronic attempts at identifying flash signatures use sophisticated sensors and discrimination techniques in both the visible and infrared regions of the spectrum. Research in this area has made significant progress, but since the electronic sensors and post processing electronics have much different sensitivities than the human eye, this research offers little information regarding human perceptions of muzzle flash and nearby objects that may be briefly illuminated by muzzle flash events.

The third main concern addressed in published muzzle flash research is whether the muzzle flash is so bright as to diminish the night vision of friendly forces during a firefight (Haag, 2007). A great deal of law enforcement and military marksmanship training and qualification occurs at well-lighted shooting ranges. However, a significant percentage of lethal force encounters occur at night. The concern is that a large fireball emerging from the muzzle of one's own firearm might disrupt the night vision of friendly forces, hindering target acquisition for some period of time during important points in the firefight. The military has specified the use of flash suppressants and flash reducing muzzle devices to ameliorate this concern, and the problem is gaining heightened awareness in the law enforcement community.

A concern that has not received much attention in published research is whether the muzzle flash of a specific cartridge provides sufficient duration and intensity of illumination in the visible region of the spectrum to allow observers to positively identify a shooter's face. The purpose of the experiment reported here is to answer this question for a specific ammunition type in a 22 LR pistol.

**Experimental Method**

The experiment used two shooters firing a total of 20 shots with two identical semi-automatic pistols chambered in 22 LR. The pistols had 6" barrels and were manufactured by Beretta. The ammunition used was CCI 22LR 40 grain lead round nose. To emphasize whether the shooter could be identified from facial features, shooters were dressed identically and used a step to make their heights comparable. Shooters fired in randomized order (generated by a spreadsheet), while eight observers (male US military veterans), attempted to identify the shooter for each trial shot.

Prior to testing, with the range well illuminated, the two shooters were identified to the observers as shooter #1 and shooter #2. The observers were given several minutes to familiarize themselves with the shooters' appearances and facial features. The range safety officer reviewed key safety instructions, and the experiment director explained the experimental procedure to the shooters, range safety officer, and observers. The observers were positioned in an area with full view of the shooters from 1-8 feet in front of the shooters and 10-15 feet to their right side. The firing range was mostly darkened with only enough ambient light to safely conduct firing tests at the direction of a range safety officer. There was not enough ambient light to identify the shooter's face. The shooters fired from a standing position using a two handed grip and making the best use of the sights as possible under the low light conditions. Each shot was taken within a few seconds of the range instructions "range is hot" and "fire when ready" so that the observers all had ample notice that a shot was about to occur. Before each shot, the experiment director announced the number of the preceeding shot and the next shot (1-20) to ensure the observers were recording their identification on the correct line of their written record. After each shot, observers attempted to identify the shooter, writing down their determination on the proper line.

Observers were all male US military veterans attending an NRA instructor course using their GI Bill benefits. Participation as observers was voluntary. Two observers reported being nearsighted, but that their nearsightedness was corrected to meet driving standards. One observer reported poor night vision. Four observers reported PTSD. Observers ranged in age from 21 to 67 with a mean age of 37.

**Results**

With 20 shots and 8 observers, the experiment had the potential for 160 possible correct identifications. Observers correctly identified the shooter 54% of the time (86/160) and failed to correctly identify the shooter 46% (74/160) of the time. Observers were unanimous that there was no hint of any visible illumination on the shooter's face provided by the muzzle flash. In cases where the shooter was correctly identified, observers reported using visual cues such as the sillhouette of the shooters' hair and slight differences in arm positions when firing the pistol. Observers reported that no identifying features were illuminated by muzzle flash. Other people in the room (including the author and an NRA training counselor positioned with the observers and the range safety officer positioned next to the shooters) were also in unanimous agreement that there was no illumination of the shooters' faces from discharge of a 22 LR pistol. When fired, the pistol produced a small yellowish flash from the end of the barrel, but that light source provided no significant or noticeable illumination of the shooter's face.

**Discussion and Conclusion**

With two shooters, random guessing would correctly identify the shooter approximately 50% of the time (80/160), just as guessing the outcome of a coin flip would tend to be right approximately 50% of the time. The result of correctly identifying the shooter 54% of the time (86/160) and failing to identify the shooter 46% (74/160) of the time is only slightly better than random guessing, and fails to support any claim that definitive identification is possible by the muzzle flash of a 22 LR pistol with the ammunution used.

Based on this experiment, it is concluded with a high degree of scientific certainty that accurate visual identification of a shooter's face is not possible from the illumination provided by a 22 LR muzzle flash. The quantity of light provided by the muzzle flash is too low and the duration of the muzzle flash is too short.

Several factors contribute to a 22 LR producing much less illumination in its muzzle flash than other cartridges. First, this 22 LR load is subsonic, which eliminates secondary ignition of unburned fuel that provides the largest contribution to muzzle flash from supersonic cartridges (including most military and law enforcement service arms). Second, 22 LR loads contain a small quantity of powder (1-2 grains). This is much less than centerfire pistol and rifle loads. For example, most 9mm NATO loads contain 5-8 grains of powder, and most 223 Remington/5.56 mm NATO loads contain 20-28 grains of powder. The low quantity of fuel, the fast powder burn rates, and the subsonic velocities of 22 LR combine to produce much smaller muzzle flash.

The small muzzle flash of the 22 LR can also be understood in terms of fundamental principles of physics. Wein's displacement law describes how the wavelength of radiation is inversely proportional to the temperature of the source. Very hot sources like the sun emit strongly in the visible spectrum (shorter wavelengths, 425-750 nm), but cooler sources like flames and combustion products emit most strongly in the infrared (IR) part of the spectrum (above 750 nm) which is not visible to the unaided human eye. Another important principle of physics describes how expanding gases rapidly cool. The small gas volume and rapid expansion of the combustion products in 22 LR cartridges cool the gases significantly before they exit the barrel. Thus, most of the thermal radiation from gases leaving the muzzle is in the IR spectrum, invisible to the human eye. (Kastek et al., 2011)

It remains an open question whether a shooter's face could be identified using night vision equipment or IR sensitive cameras or electronics. Since 22 LR muzzle flash is likely dominated by infrared radiation and since electronic detectors (including cameras) can be much more sensitive to IR than the human eye, identification with electronic methods is likely possible, though experiments testing this hypothesis would be needed for validation. (Kastek et al., 2011)